\documentclass[11pt]{article}
\title{A Note on Kuhn's Theorem\\ with Ambiguity Averse Players}
\author{Gaurab Aryal and Ronald Stauber\footnote{G. Aryal: The University of Chicago, Department of Economics, 1126 E. 59th Street, Chicago, IL 60637, Email: \texttt{aryalg@uchicago.edu}; R. Stauber (corresponding author): Research School of Economics, Australian National University, ACT 0200, Australia, Telephone: (+61) 2 6125 7138, Email: \texttt{ronald.stauber@anu.edu.au}}}
\date{June 17, 2014}
\usepackage{natbib}
\usepackage{amsfonts}
\usepackage{amssymb}
\usepackage{amsmath}
\usepackage{amsthm}
\usepackage{mathrsfs}
\usepackage{graphics}
\usepackage{color}
\usepackage{sgame}
\usepackage[left=3.2cm,right=3.2cm,top=3.2cm,bottom=3.2cm]{geometry}

\usepackage{tikz}
\usepackage{tikz-3dplot}
\usetikzlibrary{arrows}

\vspace{\bigskipamount}

\definecolor{sangre}{rgb}{0.6,0.18,0.19}
\definecolor{dullmagenta}{rgb}{0.4,0,0.4}
\definecolor{darkblue}{rgb}{0,0,0.6}

\usepackage[backref=none,pdftex,hypertexnames=false,colorlinks,urlcolor = black,citecolor = black,linkcolor=black]{hyperref}

\begin{document}

\theoremstyle{definition}
\newtheorem{defn}{Definition}
\newtheorem{ex}{Example}
\newtheorem{ass}{Assumption}

\theoremstyle{plain}
\newtheorem{thm}{Theorem}
\newtheorem{prop}{Proposition}
\newtheorem{lem}[prop]{Lemma}
\newtheorem{cor}[prop]{Corollary}

\newcommand{\eps}{\varepsilon} 
\newcommand{\marg}{\text{marg}}
\newcommand{\id}{\text{id}}
\newcommand{\cl}{\textup{cl}}
\newcommand{\co}{\textup{co}}
\newcommand{\proj}{\text{proj}}

\newcommand{\B}{\mathcal{B}}
\newcommand{\M}{\mathcal{M}}
\newcommand{\I}{\mathcal{I}}
\newcommand{\0}{\mathcal{O}}
\newcommand{\R}{\mathbb{R}}

\begin{titlepage}

\maketitle

\begin{abstract}

Kuhn's Theorem shows that extensive games with perfect recall can equivalently be analyzed using mixed or behavioral strategies, as long as players are expected utility maximizers. This note constructs an example that illustrates the limits of Kuhn's Theorem in an environment with ambiguity averse players who use a maxmin decision rule and full Bayesian updating.

\bigskip
\noindent\textit{Keywords:} Extensive games; Ambiguity; Maxmin; Dynamic consistency

\bigskip
\noindent\textit{JEL Classification:} C72; D81
\end{abstract}

\thispagestyle{empty}

\end{titlepage}


\section{Introduction}

A classic result in game theory, Kuhn's Theorem \citep{kuh53} shows that for each mixed strategy in an extensive game with perfect recall, there exists an outcome-equivalent behavioral strategy, and vice versa.\footnote{See \citet{msz13} for a thorough exposition of this result. Two strategies are outcome-equivalent if they induce the same distribution over terminal histories, irrespective of the strategies used by a player's opponents.} In light of the fact that expected utility maximizers are dynamically consistent based on Bayesian updating, the principal implication of Kuhn's Theorem is that the optimality of a mixed strategy in the strategic form of an extensive game is equivalent to the conditional optimality of its outcome-equivalent behavioral strategy, at all information sets that are reached with positive probability according to a player's beliefs about the strategies of his opponents, as pointed out by \citet{bra07}. 

This equivalence between mixed and behavioral strategies does not extend to games where players are ambiguity averse in the sense of \citet{gisc89}, so that beliefs are represented by sets of probability measures, and a maxmin decision rule is used to define optimality. Such set-valued beliefs of a player in an extensive game may arise in a number of settings, which could include Bayesian games where players have ambiguous information about opponents' types as in \citet{kaui05}, games where equilibria are defined by ambiguous beliefs as in \citet{lo96}, games where players are allowed to use ambiguous randomization devices as in \citet{risa13}, or games where players' beliefs are defined through ambiguous trembles of their opponents' unambiguous strategies as in \citet{arst13}. If we consider an extensive game with complete (but not necessarily perfect) information, so that each player $i$ only faces uncertainty regarding his opponents' strategies, and this uncertainty is represented by set-valued beliefs, the standard Kuhn's Theorem still yields the existence of (unambiguous) behavioral and mixed strategies $\beta_i$ and $\mu_i$ for this player that are outcome-equivalent for every element of his (set-valued) beliefs. However, if this player uses full Bayesian updating to derive beliefs at each information set, combined with a maxmin decision rule, optimality of the mixed strategy $\mu_i$ in the strategic form need not correspond to the conditional optimality of the outcome-equivalent behavioral strategy $\beta_i$ at information sets that are reached according to any element of the player's ambiguous beliefs. This is a straightforward consequence of the well-known fact that maxmin expected utility is not dynamically consistent with full Bayesian updating.\footnote{See, for example, \citet{epsc03} and \citet{eple93}.} 

In an interesting paper, \citet{epsc03} show that dynamic consistency can be restored in dynamic choice settings with maxmin preferences and full Bayesian updating, as long as the overall uncertainty a decision-maker faces can be captured by a set of priors that has a ``rectangularity" property. Roughly, rectangularity means that the initial set of priors can be constructed by recursively combining all its corresponding conditional and marginal probabilities at each stage of the information filtration. \citet{sas13} shows that the characterization of \citet{epsc03} can be applied in extensive games with perfect information to recover a version of Kuhn's Theorem, 
which implies that in environments with perfect information, the dynamic consistency associated to the standard Kuhn's Theorem can be recovered through the use of mixed strategies that possess a type of rectangularity property. The principal contribution of this note is to consider whether this approach can be extended to derive restrictions on ex-ante ambiguous beliefs/strategies, such as the rectangularity property identified by \citet{epsc03}, that yield dynamic consistency in extensive games with imperfect information. We show by means of an example that such restrictions are not possible, \emph{if} any two players' ambiguous beliefs about a third opponent's strategies are required to be consistent---in the sense that these beliefs are the same---and each player's beliefs regarding the strategies of his opponents are independent across players.

\section{Analysis}

We first present an example that shows that the dynamic consistency associated to standard expected utility maximization may not hold in extensive games with ambiguity averse players, which implies that an analysis based on the extensive form of a game may yield different conclusions than one based on its strategic form. We then show how dynamic consistency might be recovered in this example, using the rectangularity property identified by \citet{epsc03}. Finally, we construct an extension of the example which shows that this approach to achieving dynamic consistency cannot be extended to all extensive games with perfect recall. Our examples are all standard extensive games with imperfect information and perfect recall, as defined in \citet{msz13}, for example.

\bigskip

\begin{figure}
\begin{center}
\begin{tikzpicture}
\begin{scope}[semithick]

\draw (6.3,4.5)  -- node[above left=-2pt]{$L$} (1,3);
\draw (6.3,4.5)  -- node[above left=-3pt]{$R$} (5,3);
\draw (6.3,4.5)  -- node[above right=-2pt]{$O$} (9,1.5) node[below] {$x,-1$};

\filldraw[fill=white,draw=black] (6.3,4.5) node[above] {$1$} circle (2pt);

\draw (1,3)  -- node[left]{$M$} (0,1.5) node[below] {$x,0$};
\draw (1,3)  -- node[right]{$N$} (2,1.5) node[below] {$x,101$};

\draw (5,3)  -- node[left]{$M$} (4,1.5) node[below] {$x,101$};
\draw (5,3)  -- node[right]{$N$} (6,1.5) node[below] {$x,100$};

\filldraw[black] (1,3)  circle (2pt);
\filldraw[black] (5,3)  circle (2pt);

\draw[loosely dashed](1,3) -- node[above right=-2pt]{$2$} (5,3);

\end{scope}
\end{tikzpicture}
\end{center}
\caption[]{A two-player example.}\label{ex1}
\end{figure}
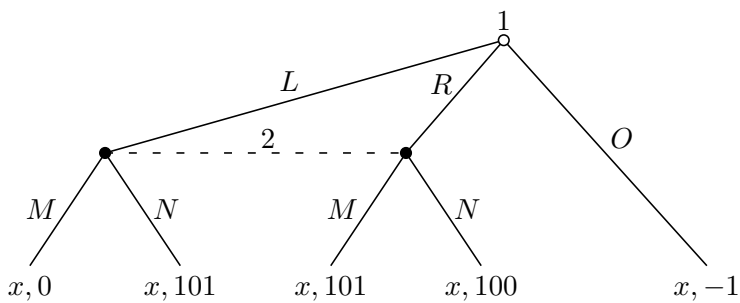

Consider the game described in Figure \ref{ex1}, which is a slight modification of an example from \citet{arst13}. Since each player only moves at one information set, the two players' sets of mixed and behavioral strategies are identical, so Kuhn's Theorem holds trivially. However, if we interpret mixed strategies as strategies in the strategic form associated to the extensive game, choosing a mixed strategy by player 2 implicitly requires that this player commits to a randomization over actions before the start of the game, and hence before he knows whether his information set is reached or not, following player 1's actions. If 2's beliefs about 1's strategies are unambiguous, and if 2's information set is reached with strictly positive probability according to these beliefs, the ex ante optimality of a (mixed) strategy of player 2 is equivalent to the conditional optimality of the respective (behavioral) strategy at 2's information set. This is a consequence of the dynamic consistency of expected utility maximization.

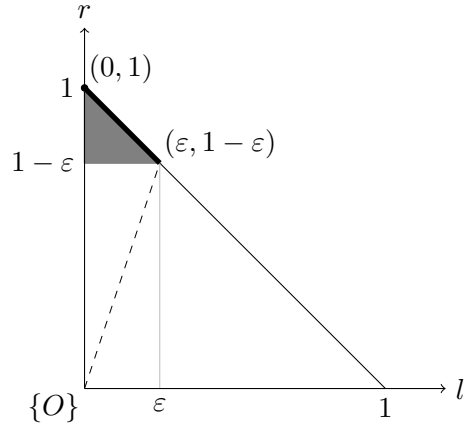
\begin{figure}
\begin{center}
\begin{tikzpicture}[scale=4]

\draw[semithick,fill=gray,gray] (0,3/4) -- (1/4,3/4) --(0,1) --(0,3/4);

\filldraw[black] (0,1)  circle (0.3pt);

\draw (1,0) node[below] {$1$} -- (0,1) node[left]{$1$};

\node [left] at (0,3/4)  {$1-\eps$};
\draw[thin, lightgray] (1/4,3/4) -- (1/4,0);
\node [below] at (1/4,0) {$\eps$};

\draw[very thin, dashed] (0,0) -- (1/4,3/4);

\draw[line width=2] (0,1) node[above right=-3pt] {$\left(0,1\right)$} -- (1/4,3/4) node[above right=-3pt]{$\left(\eps,1-\eps\right)$};

\draw [->] (0,0)  --  (1.2,0) node[right] {$l$};
\draw [->] (0,0)  --  (0,1.2) node[above] {$r$};
\node [below left=-2pt] at (0,0) {$\{O\}$};
\end{tikzpicture}
\end{center}
\caption[]{An $\eps$-contamination.}\label{cont}
\end{figure}

Now assume that player 2 is ambiguity averse, and that his (ambiguous) beliefs about player 1's strategy are given by an $\eps$-contamination of the strategy that assigns probability 1 to action $R$.\footnote{See \citet{arst13} for an interpretation of such beliefs.} Denoting the strategy assigning probability 1 to $R$ by $\beta_1^*=(l^*,r^*,o^*)=(0,1,0)$, and the set of all distributions over $\{L,R,O\}$ by $\Delta_1$, an $\eps$-contamination of $\beta_1^*$ for some small $\eps>0$ is defined by the set of distributions
\[\beta_1^\eps:=(1-\eps)\beta_1^*+\eps \Delta_1.\]
The projection of the set $\beta_1^\eps$ onto the $l$-$r$ plane is illustrated by the gray-shaded triangle in Figure \ref{cont}.\footnote{The origin of the coordinate system corresponds to player 1 playing $O$ with probability 1.} Denote a strategy of player 2 by $\beta_2=(m,1-m)$, where $m$ is the probability 
assigned to action $M$, and assume that player 2 chooses a value of $m$ in the strategic form of the game, and hence is able to commit to $m$ before knowing whether his information set is reached or not. If 2 follows a maxmin decision rule given $\beta_1^\eps$, he then chooses $m$ to solve
\[\max_{m\in[0,1]} \min_{(l,r,o)\in \beta_1^\eps} 
\{o(-1)+l[101(1-m)]+r[101m+100(1-m)]\}.\]
Clearly, for every $m\in[0,1]$, the minimum in this problem is attained at $(l,r,o)=(0,1-\eps,\eps)$, and hence 2's optimal choice is to set $m=1$.

In order to consider 2's optimal strategy conditional on reaching his information set, we derive his conditional beliefs at the information set using full Bayesian updating of all distributions in $\beta_1^\eps$.\footnote{Note that in this particular case, 2's information set is reached with positive probability according to every element of $\beta_1^\eps$.} In Figure \ref{cont}, the resulting conditional beliefs over the information set $\{L,R\}$ are represented by the thick black line connecting the points $(0,1)$ and $(\eps,1-\eps)$. Letting $\delta$ denote the probability assigned to history $R$, 2's optimal choice of a (behavioral) strategy must solve
\begin{align*}
&\max_{m\in[0,1]}\min_{\delta\in[1-\eps,1]}\{\delta[101m+100(1-m)]+(1-\delta)(1-m)101\}\\
&=\max_{m\in[0,1]}\min_{\delta\in[1-\eps,1]}\{\delta(102m-1)+101(1-m)\}\\
&=\max_{m\in[0,1]}\begin{cases}100+m, &\text{if}\,\,\, m\leq\frac{1}{102},\\
100+\eps +(1-102\eps)m, &\text{if}\,\,\, m>\frac{1}{102}.
\end{cases}
\end{align*}
It follows that as long as $1-102\eps <0$, or equivalently, $\eps>\frac{1}{102}$, the optimal conditional strategy of player 2 is to set $m=\frac{1}{102}$, so his optimal choice is not dynamically consistent.

\begin{figure}
\begin{center}
\begin{tikzpicture}[scale=4]

\draw[semithick,fill=gray,gray] (0,3/4) -- (1/4,3/4) --(0,1) --(0,3/4);

\draw[semithick,fill=lightgray,lightgray] (0,3/4) -- (3/16,9/16) -- (1/4,3/4)  --(0,3/4);

\filldraw[black] (0,1)  circle (0.3pt);

\draw (1,0) node[below] {$1$} -- (0,1) node[left]{$1$};

\node [left] at (0,3/4)  {$1-\eps$};
\draw[thin, lightgray] (1/4,3/4) -- (1/4,0);
\node [below] at (1/4,0) {$\eps$};

\draw[very thin, dashed] (0,0) -- (1/4,3/4);

\draw[line width=2] (0,1) node[above right=-3pt] {$\left(0,1\right)$} -- (1/4,3/4) node[above right=-3pt]{$\left(\eps,1-\eps\right)$};

\draw[thin, lightgray] (0,3/4) -- (3/4,0);

\draw [->] (0,0)  --  (1.2,0) node[right] {$l$};
\draw [->] (0,0)  --  (0,1.2) node[above] {$r$};
\node [below] at (3/4,0) {$1-\eps$};
\node [below left=-2pt] at (0,0) {$\{O\}$};
\end{tikzpicture}
\end{center}
\caption[]{The smallest rectangular set of priors containing $\beta_1^\eps$.}\label{rect}
\end{figure}
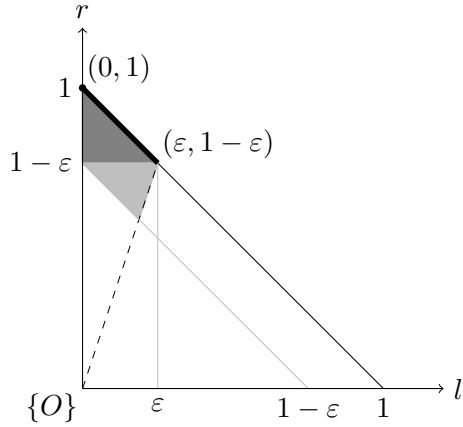

If we interpret 2's beliefs as defined by the $\eps$-contamination as a set of priors corresponding to a dynamic decision problem for player 2, then this set of priors does not satisfy the rectangularity condition of \citet{epsc03} that guarantees dynamic consistency, so the dynamic inconsistency is not unexpected. We now explain the implications of rectangularity for player 2's dynamic decision problem, referring to \citet{epsc03} for a general definition. 2's information structure is given by a filtration $(\mathscr{F}_2^0,\mathscr{F}_2^1)$, where $\mathscr{F}_2^0=\{L,R,O\}$, $\mathscr{F}_2^1=\{\{L,R\},\{O\}\}$, and all information is revealed after 2 chooses an action. The restriction of every prior $p^0\in \Delta_1$ to $\mathscr{F}_2^1$ defines a marginal, or \emph{one-step-ahead}, distribution $p^0_+$ over $\mathscr{F}_2^1$, such that $p^0$ can be decomposed in terms of $p^0_+$ and its conditional given $\mathscr{F}_2^1$, denoted by $p^1$, using the standard form
\[p^0=\int_{\mathscr{F}_2^0}p^1dp^0_+.\]
A set of priors $P\in \Delta_1$ is \emph{rectangular} if it is defined by all the compositions of all its conditionals and one-step-ahead distributions, so for every one-step-ahead distribution $p^0_+$ corresponding to some $p^0\in P$, and every conditional $q^1$ corresponding to some $q^0\in P$, the composition $ \int_{\mathscr{F}_2^0}q^1dp^0_+ $ is also an element of $P$. \citet{epsc03} note that any set of conditionals $P^1$ combined with any set of one-step-ahead distributions $P^0_+$ define an induced rectangular set of priors given by 
\[\mathscr{P}(P^1,P^0_+):=\left\{\left. \int_{\mathscr{F}_2^0} q^1dp^0_+\,\right|\, q^1\in P^1,\,\,\,p^0_+\in P^0_+\right\}.\]
In our example, player 2's beliefs are defined by a set of priors $P\in \Delta_1$ such that $P=\beta_1^\eps$. If $P^1$ and $P^0_+$ denotes the sets of all conditionals and one-step-ahead distributions associated to elements of $P$, then $\mathscr{P}(P^1,P^0_+)$ is the smallest rectangular set of priors containing $P$, as noted by \citet{epsc03}. The set of induced one-step-ahead distributions is given by all distributions that assign a probability between $0$ and $\eps$ to the event $\{O\}$, and the remaining probability to the event $\{L,R\}$. Combining these distributions with the previously described conditionals according to the formula for $\mathscr{P}(P^1,P^0_+)$, yields the smallest rectangular set of priors containing $P=\beta_1^\eps$. In Figure \ref{rect}, the rectangular set $\mathscr{P}(P^1,P^0_+)$ is given by the union of the gray and light gray shaded triangles. As explained in \citet{epsc03}, the two lines with slope $-1$ form the boundary of the region where the probability of $\{O\}$ lies between $0$ and $\eps$, and the vertical axis together with the dashed line form the boundary of the region that induces the conditionals over $\{L,R\}$ represented by the thick black line. Combining the associated one-step-ahead and conditional distributions yields the resulting rectangular set.

If we now consider player 2's problem resulting from (ambiguous) beliefs induced by the rectangular set $\mathscr{P}(P^1,P^0_+)$, we get the same optimal choice of $m$ at 2's information set, since the set of conditional beliefs resulting from full Bayesian updating does not change. However, the ex ante problem of choosing a mixed strategy in the strategic form now becomes
\[\max_{m\in[0,1]} \min_{p_+\in [0,\eps],\,\delta\in[1-\eps,1]}
\left\{p_+(-1) +(1-p_+)\left\{\delta[101m+100(1-m)]+(1-\delta)(1-m)101
\right\}\right\},\]
which clearly has the same solution as the conditional problem, and hence dynamic consistency is restored.

\bigskip

The previous example suggests that it might be possible to restore dynamic consistency in extensive games by restricting every player's beliefs about his opponents' strategies in a way that generates rectangular priors for the filtration defining the player's associated decision problem. We show next, by constructing an appropriate example, that this approach cannot be used to ensure dynamic consistency in general games.

\bigskip

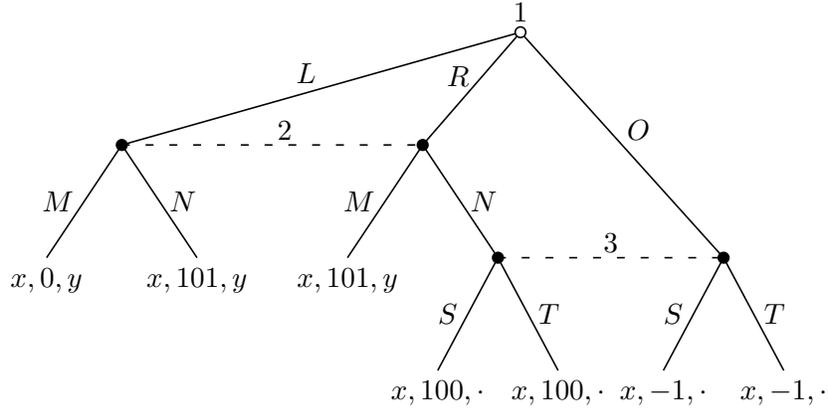
\begin{figure}
\begin{center}
\begin{tikzpicture}
\begin{scope}[semithick]

\draw (6.3,4.5)  -- node[above left=-2pt]{$L$} (1,3);
\draw (6.3,4.5)  -- node[above left=-3pt]{$R$} (5,3);
\draw (6.3,4.5)  -- node[above right=-2pt]{$O$} (9,1.5);

\filldraw[fill=white,draw=black] (6.3,4.5) node[above] {$1$} circle (2pt);

\draw (1,3)  -- node[left]{$M$} (0,1.5) node[below] {$x,0,y$};
\draw (1,3)  -- node[right]{$N$} (2,1.5) node[below] {$x,101,y$};

\draw (5,3)  -- node[left]{$M$} (4,1.5) node[below] {$x,101,y$};
\draw (5,3)  -- node[right]{$N$} (6,1.5);

\filldraw[black] (1,3)  circle (2pt);
\filldraw[black] (5,3)  circle (2pt);

\draw[loosely dashed](1,3) -- node[above right=-2pt]{$2$} (5,3);

\draw (6,1.5)  -- node[left]{$S$} (5.2,0) node[below] {$x,100,\cdot$};
\draw (6,1.5)  -- node[right]{$T$} (6.8,0) node[below] {$x,100,\cdot$};

\draw (9,1.5)  -- node[left]{$S$} (8.2,0) node[below] {$x,-1,\cdot$};
\draw (9,1.5)  -- node[right]{$T$} (9.8,0) node[below] {$x,-1,\cdot$};

\filldraw[black] (6,1.5)  circle (2pt);
\filldraw[black] (9,1.5)  circle (2pt);

\draw[loosely dashed](6,1.5) -- node[above=-2pt]{$3$} (9,1.5);
\end{scope}
\end{tikzpicture}
\end{center}
\caption[]{A three-player example.}\label{ex2}
\end{figure}

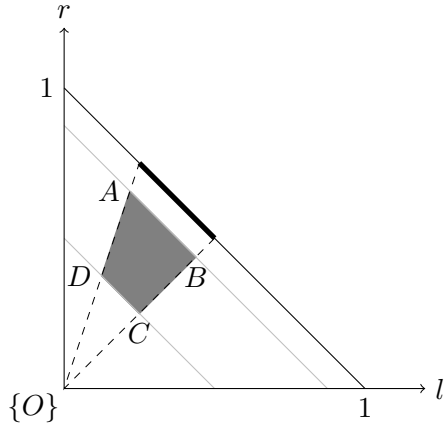
\begin{figure}
\begin{center}
\begin{tikzpicture}[scale=4]

\draw[semithick,fill=gray,gray] (7/32,21/32) -- (7/16,7/16) --(1/4,1/4) --(1/8,3/8) -- (7/32,21/32);

\draw (1,0) node[below] {$1$} -- (0,1) node[left]{$1$};

\draw[thin, lightgray] (0,7/8) -- (7/8,0);
\draw[thin, lightgray] (0,1/2) -- (1/2,0);

\draw[very thin, dashed] (1/2,1/2) -- (0,0) -- (1/4,3/4);
\draw[line width=2] (1/4,3/4) -- (1/2,1/2);

\node [left] at (7/32,21/32) {$A$};
\node [below] at (7/16,7/16) {$B$};

\node [below] at (1/4,1/4) {$C$};
\node [left] at (1/8,3/8) {$D$};

\draw [->] (0,0)  --  (1.2,0) node[right] {$l$};
\draw [->] (0,0)  --  (0,1.2) node[above] {$r$};
\node [below left=-2pt] at (0,0) {$\{O\}$};
\end{tikzpicture}
\end{center}
\caption[]{A rectangular set of priors for player $2$.}\label{rec2}
\end{figure}

Consider the three player game described in Figure \ref{ex2}. This game is an extension of the two player game we analyzed previously, and since player 2's payoffs are assumed to be independent of player 3's actions, our previous analysis and the result of \citet{epsc03}, which shows the equivalence of dynamic consistency and rectangular priors, imply that in order to guarantee dynamic consistency for player 2, we need to assume that 2's beliefs about 1's actions are given by a rectangular prior. Since 2's information is given by the same filtration as in the previous game, any rectangular set of priors must be of the form described by the gray area in Figure \ref{rec2}. If player 3's (ambiguous) beliefs about player 1's actions are required to be consistent with player 2's beliefs, then the rectangularity requirements that we need to ensure dynamic consistency for player 2, imply that 3's beliefs about 1's actions must be rectangular relative to 2's information filtration. We thus assume that players 2 and 3 have identical priors over 1's actions, given by a rectangular set such as the one described in Figure \ref{rec2}. Furthermore, assuming that 3's beliefs about 2's actions are independent of his beliefs about 1's actions, we can represent such (ambiguous) beliefs by an interval $[a,b]\subseteq [0,1]$, where any $n\in [a,b]$ denotes the probability 2 assigns to action $N$.

\begin{figure}
\begin{center}
\begin{tikzpicture}[scale=4]

\draw (1,0) node[below] {$1$} -- (0,1) node[left]{$1$};

\draw[thin, lightgray] (0,3/4) -- (3/4,0);
\draw[thin, lightgray] (0,9/16) -- (3/16,9/16) -- (3/16,0);
\draw[thin, lightgray] (0,9/32) -- (15/32,9/32) -- (15/32,0);

\node [left] at (0,3/4)  {$1-o$};
\node [below] at (3/4,0) {$1-o$};

\node [left] at (0,9/16)  {$r$};
\node [below] at (3/16,0) {$l$};

\node [left] at (0,9/32)  {$rn$};
\node [below] at (15/32,0) {$z$};

\filldraw[black] (3/16,9/16)  circle (0.4pt);
\filldraw[black] (15/32,9/32)  circle (0.4pt);

\node [above right=-2pt] at (3/16,9/16) {$E$};
\node [above right=-2pt] at (15/32,9/32) {$F$};

\draw [very thick,->] (3/16,9/16) -- (15/32,9/32);

\draw [->] (0,0)  --  (1.2,0) node[right] {$l,z$};
\draw [->] (0,0)  --  (0,1.2) node[above] {$r, rn$};
\node [below left=-2pt] at (0,0) {$\{O\}$};
\end{tikzpicture}
\end{center}
\caption[]{Changing priors.}\label{chg}
\end{figure}
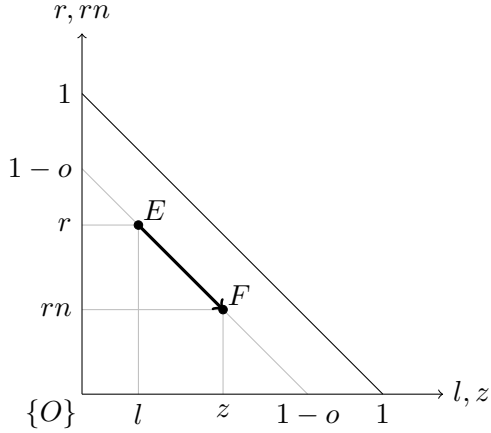

To analyze dynamic consistency for player 3, note that his information filtration is given by $\mathscr{F}_3^0=\{LM, LN, RM, RN, O\}$ and $\mathscr{F}_3^1=\{\{LM, LN, RM\}, \{RN, O\}\}$. Since 3's payoffs arising from the terminal histories $LM$, $LN$ and $RM$ are assumed to be identical, we can combine these histories into a single ``state" for player 3, which we will denote by $Z$. We can then rewrite the filtration as $\mathscr{F}_3^0=\{Z, RN, O\}$ and $\mathscr{F}_3^1=\{\{Z\}, \{RN, O\}\}$, and represent any prior in a two-dimensional ``Machina triangle" where the horizontal axis represents the probability of state $Z$, denoted by $z$, the vertical axis represents the probability of state $RN$, denoted by $rn$, and the origin is equivalent to state $O$. Given a strategy of player 1 denoted by $(l,r,o)$, and an independent strategy of player 2 denoted by $n$, the induced prior over 3's state space assigns probability $o$ to state $O$, probability $rn$ to state $RN$, and probability $l+r(1-n)$ to state $Z$. Since the origins of the triangles representing the prior $(l,r,o)$ for players 2 and the corresponding prior $(l+r(1-n), rn,o)\equiv (z,rn,o)$ for player 3 both coincide with probability 1 assigned to state $O$, we can represent both priors on the same triangle. As the probability of $O$ stays constant, a ``shift" from $(l,r,o)$ to $ (z,rn,o)$ is captured by a movement along the line of slope $-1$ that corresponds to the probability $o$, as described by a move from point $E$ to point $F$ in Figure \ref{chg}.

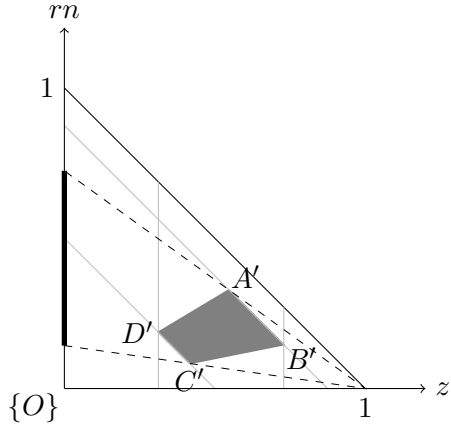
\begin{figure}
\begin{center}
\begin{tikzpicture}[scale=4]

\draw[semithick,fill=gray,gray] (35/64,21/64) -- (35/48,7/48) --(5/12,1/12) --(5/16,3/16) -- (35/64,21/64);

\draw (1,0) node[below] {$1$} -- (0,1) node[left]{$1$};

\draw[thin, lightgray] (0,7/8) -- (7/8,0);
\draw[thin, lightgray] (0,1/2) -- (1/2,0);

\draw[thin, lightgray] (35/48,0) -- (35/48,13/48);
\draw[thin, lightgray] (5/16,0) -- (5/16,11/16);

\draw[very thin, dashed] (0,1/7) -- (1,0) -- (0,21/29);
\draw[line width=2] (0,1/7) -- (0,21/29);

\node [above right=-3pt] at (35/64,21/64)  {$A'$};
\node [below right=-3pt] at (35/48,7/48) {$B'$};

\node [below=-2pt] at (5/12,1/12) {$C'$};
\node [left=-2pt] at (5/16,3/16) {$D'$};

\draw [->] (0,0)  --  (1.2,0) node[right] {$z$};
\draw [->] (0,0)  --  (0,1.2) node[above] {$rn$};
\node [below left=-2pt] at (0,0) {$\{O\}$};
\end{tikzpicture}
\end{center}
\caption[]{Induced set of priors for player $3$.}\label{ind3}
\end{figure}

If both 2's and 3's beliefs about 1's actions are captured by the same arbitrary set of priors that is rectangular with respect to 2's information filtration, such as the one described in Figure \ref{rec2}, and 3's beliefs about 2's actions are captured by a set $[a,b]$ representing probabilities $n$ assigned to action $N$, we can represent the induced set of priors for player 3's filtration by shifting all points of the rectangular set spanned by the points $\{A,B,C,D\}$ in Figure \ref{rec2}, analogously to the shift from Figure \ref{chg}. As long as $0<a<b<1$, this would imply that each point $(l,r,o)$ from the initial rectangular set will yield a set of points along the line of slope $-1$ corresponding to $o$, such that the corresponding set of ``$rn$-coordinates'' lie in the interval $[ra,rb]$. The full initial set spanned by $\{A,B,C,D\}$ in the $l$-$r$ plane then yields a set of priors for player 3 in the $z$-$rn$ plane, which is represented in Figure \ref{ind3} by the gray shaded area spanned by $\{A',B',C',D'\}$. Note that the northwest boundary $A'D'$ of this induced set of priors corresponds to the northwest boundary $AD$ of the initial set of priors, transformed using the larger value $n=b$; similarly, the southeast boundary $B'C'$ is induced by the initial southeast boundary $BC$, transformed using the smaller value $n=a$.\footnote{Note that any point $( \alpha l + (1-\alpha) l', \alpha r +(1-\alpha) r')$ on the line connecting two elements $(l,r)$ and $(l',r')$ of the initial rectangular set, if transformed using some value of $n$, will lie on the line connecting the respective transformations of  $(l,r)$ and $(l',r')$, if based on the same value of $n$, since
\[(\alpha l + (1-\alpha) l' + [\alpha r +(1-\alpha) r'](1-n), [\alpha r +(1-\alpha) r']n)=(\alpha[l+r(1-n)]+(1-\alpha)[l'+r'(1-n)], \alpha rn+(1-\alpha)r'n).\]
This implies that the boundaries $A'D'$ and $B'C'$ must always be straight lines. Furthermore, since the initial points $A$ and $D$ yield identical conditionals over $\{L,R\}$, the points $A'$ and $D'$ will yield identical conditionals over $\{Z,RN\}$, and similarly for $B'$ and $C'$.} 
Hence, if $r_A$ and $r_D$, and $l_A$ and $l_D$ denote the $r$- and $l$-coordinates, respectively, of the points $A$ and $D$ in Figure \ref{rec2}, then the $rn$-coordinates of the points $A'$ and $D'$ in Figure \ref{ind3} will be given by $r_Ab$ and $r_Db$, and the $z$-coordinates of $A'$ and $D'$ are $z_{A'}=l_A+r_A(1-b)$ and $z_{D'}=l_D+r_D(1-b)$. Since $r_A>r_D$ and $l_A>l_D$, this implies that both the $rn$ and the $z$ coordinates of $A'$ exceed those of $D'$. A similar argument shows that the $rn$ and the $z$ coordinates of $B'$ exceed those of $C'$. Since the lines connecting $A'$ and $B'$, and $C'$ and $D'$, must both have slope $-1$, this implies that the set spanned by $\{A',B',C',D'\}$ cannot be rectangular relative to player 3's filtration, because such a rectangular set would have to be bounded by two vertical lines representing constant values of $z$, and by two lines that bound the set of all priors that map to some interval of conditional beliefs over $\{RN,O\}$ (such as the thick black line on the $rn$-axis in Figure \ref{ind3}). The failure of rectangularity for player 3's information filtration implies that there exist payoffs for player 3 under which dynamic consistency does not hold.

\section{Conclusion}

Our analysis shows that if any two players' ambiguous beliefs about a third player in a game are required to be consistent, and each player's beliefs regarding the strategies of his opponents are assumed to be independent across players, then the dynamic consistency implicit in Kuhn's Theorem cannot be achieved in every extensive game with imperfect information and ambiguity averse players, even if additional rectangularity assumptions are introduced. The obvious implication is that an analysis of such games must rely on the extensive form and on a consistent planning assumption in the spirit of \citet{str55}. Alternatively, if an analyst wants to use the strategic form of a game and retain dynamic consistency, he may do so by using ambiguous beliefs that are rectangular, but then he must either allow for inconsistencies of beliefs across players, or restrict his analysis to extensive games forms for which rectangular beliefs are guaranteed to yield dynamic consistency. A recent paper by \citet*{mrs14} identifies restrictions that characterize a corresponding class of games.


{\small\bibliographystyle{elsarticle-harv}
\bibliography{kuhn}}

\end{document}